\documentclass[10pt,twocolumn,letterpaper]{article}

\usepackage{cvpr}
\usepackage{times}
\usepackage{epsfig}
\usepackage{graphicx}
\usepackage{amsmath}
\usepackage{amssymb}

\usepackage[breaklinks=true,bookmarks=false]{hyperref}

\cvprfinalcopy 

\def\cvprPaperID{10} 

\begin{document}

\title{Deep Learning for Automatic Pneumonia Detection}

\author{Tatiana Gabruseva\\
Independent researcher\\
tatigabru.com\\
{\tt\small tatigabru@gmail.com}
\and
Dmytro Poplavskiy \\
Topcon Positioning Systems\\
Brisbane, Queensland, Australia\\
\and
Alexandr A. Kalinin \\
University of Michigan\\
Ann Arbor, MI 48109 USA, and\\
Shenzhen Research Institute of Big Data, \\
Shenzhen 518172, Guangdong, China\\
{\tt\small akalinin@umich.edu}
}

\maketitle

\begin{abstract}
Pneumonia is the leading cause of death among young children and one of the top mortality causes worldwide. The pneumonia detection is usually performed through examine of chest X-ray radiograph by highly-trained specialists. This process is tedious and often leads to a disagreement between radiologists. Computer-aided diagnosis systems showed the potential for improving diagnostic accuracy. In this work, we develop the computational approach for pneumonia regions detection based on single-shot detectors, squeeze-and-excitation deep convolution neural networks, augmentations and multi-task learning. The proposed approach was evaluated in the context of the Radiological Society of North America Pneumonia Detection Challenge, achieving one of the best results in the challenge.

\end{abstract}

\section{Introduction}
Pneumonia accounts for around 16\% of all deaths of children under five years worldwide~\cite{who_2018}, being the world’s leading cause of death among young children~\cite{pneumonia_facts}. In the United States only, about $1$ million adults seek care in a hospital due to pneumonia every year, and $50,000$ die from this disease~\cite{pneumonia_facts}. The pneumonia complicating recent coronavirus disease 2019 (COVID-19) is a life-threatening condition claiming thousands of lives in 2020~\cite{Duployez2020, Fu2020,COVID19}. Pneumonia caused by COVID-19 is of huge global concern, with confirmed cases in 185 countries across five continents at the time of writing this paper~\cite{COVID19}.

The pneumonia detection is commonly performed through examine of chest X-ray radiograph (CXR) by highly-trained specialists. It usually manifests as an area or areas of increased opacity on CXR~\cite{Franquet2018}, the diagnosis is further confirmed through clinical history, vital signs and laboratory exams. The diagnosis of pneumonia on CXR is complicated due to the presence of other conditions in the lungs, such as fluid overload, bleeding, volume loss, lung cancer, post-radiation or surgical changes. When available, comparison of CXRs of the patient taken at different time points and correlation with clinical symptoms and history is helpful in making the diagnosis. A number of factors such as positioning of the patient and depth of inspiration can alter the appearance of the CXR~\cite{Kelly_2012}, complicating interpretation even further.

There is a known variability between radiologists in the interpretation of chest radiographs~\cite{Neuman2011}. To improve the efficiency and accuracy of diagnostic services computer-aided diagnosis systems for pneumonia detection has been widely exploited in the last decade~\cite{Oliveira2008,Noor2010,Shabnam2011,Zech2018,chexnet_2017}. Deep learning approaches outperformed conventional machine learning methods in many medical image analysis tasks, including detection~\cite{chexnet_2017}, classification~\cite{Rakhlin2018} and segmentation~\cite{Ronneberger2015,kalinin2020medical}. Here, we present the solution of the Radiological Society of North America (RSNA) Pneumonia Detection Challenge for pneumonia regions detection hosted on Kaggle platform~\cite{kaggle_overview}. Our approach uses a single-shot detector (SSD), squeeze-and-excitation deep convolution neural networks (CNNs)~\cite{Hu_2018_CVPR}, augmentations and multi-task learning. The algorithm automatically locates lung opacities on chest radiographs and demonstrated one of the best performance in the challenge. The source code is available at https://github.com/tatigabru/kaggle-rsna.

\section{Dataset}
The labelled dataset of the chest X-ray images and patients metadata was publicly provided for the challenge by the US National Institutes of Health Clinical Center~\cite{dataset_2017}. This database comprises frontal-view X-ray images from $26684$ unique patients. Each image was labelled with one of three different classes from the associated radiological reports: "Normal",  "No Lung Opacity / Not Normal", "Lung Opacity".

Usually, the lungs are full of air. When someone has pneumonia, the air in the lungs is replaced by other material, i.e. fluids, bacteria, immune system cells, etc. The lung opacities refers to the areas that preferentially attenuate the X-ray beam and therefore appear more opaque on CXR than they should, indicating that the lung tissue in that area is probably not healthy.

The "Normal" class contained data of healthy patients without any pathologies found (including, but not limited to pneumonia, pneumothorax, atelectasis, etc.). The "Lung Opacity" class had images with the presence of fuzzy clouds of white in the lungs, associated with pneumonia. The regions of lung opacities were labelled with bounding boxes. Any given patient could have multiple boxes if more than one area with pneumonia was detected. There are different kinds of lung opacities, some are related to pneumonia and some are not. The class "No Lung Opacity / Not Normal" illustrated data for patients with visible on CXR lung opacity regions, but without diagnosed pneumonia.
\begin{figure*}[htb]
\begin{center}
    \includegraphics[width=0.9\linewidth]{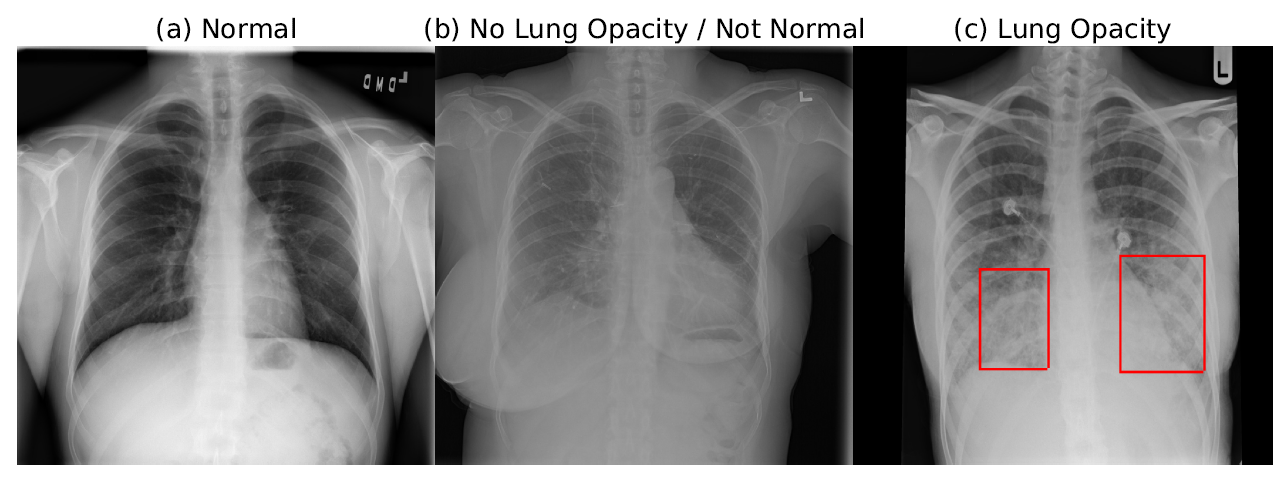}
\end{center}
   \caption{Examples of the chest X-ray images for (a) "Normal",  (b) "No Lung Opacity / Not Normal", and (c) "Lung Opacity" cases.  The lung opacities regions are shown on (c) with red bounding boxes.}
\label{fig:eda}
\end{figure*}
Fig.~\ref{fig:eda} shows examples of CXRs for all three classes labeled with bounding boxes for unhealthy patients.

The dataset is well-balanced with the distribution of classes as shown in Table 1.
\begin{table}[htb]
\begin{center}
\begin{tabular}{|l|c|c|}
\hline
Class & Target & Patients \\
\hline\hline
Lung Opacity & 1 & 9555\\
No Lung Opacity / Not Normal & 0 & 11821 \\
Normal & 0 & 8851\\
\hline
\end{tabular}
\end{center}
\label{table:classes}
\caption{Classes distribution in the dataset. Target 1 or 0 indicates weather pneumonia is diagnosed or not, respectively.}
\end{table}

\section{Evaluation}
Models were evaluated using the mean average precision (mAP) at different intersection-over-union (IoU) thresholds~\cite{evaluation}. The threshold values ranged from 0.4 to 0.75 with a step size of 0.05: (0.4, 0.45, 0.5, 0.55, 0.6, 0.65, 0.7, 0.75). A predicted object was considered a "hit" if its intersection over union with a ground truth object was greater than 0.4. The average precision ($AP$) for a single image was calculated as the mean of the precision values at each IoU threshold as following:
\begin{equation}
    AP = \frac{1}{|thresholds|} \sum_t \frac{TP(t)}{TP(t) + FP(t) + FN(t)}
\end{equation}
Lastly, the score returned by the competition metric, $mAP$, was the mean taken over the individual average precisions of each image in the test dataset.

\section{Model}
Often, the solutions in machine learning competitions are based on large and diverse ensembles, test-time augmentation, and pseudo labelling, which is not always possible and feasible in real-life applications. At test-time, we often want to minimize a memory footprint and inference time. Here, we propose a solution based on a single model, ensembled over several checkpoints and $4$ folds. The model utilises an SSD RetinaNet~\cite{retina} with SE-ResNext101 encoder pre-trained on ImageNet~\cite{imagenet}.

\subsection{Base model}
The model is based on RetinaNet~\cite{retina} implementation on Pytorch~\cite{paszke2019pytorch} with the following modifications:
\begin{enumerate}
	\item Images with empty boxes were added to the model and contributed to the loss calculation/optimisation (the original Pytorch RetinaNet implementation~\cite{retinanet} ignored images with no boxes).
	\item The extra output for small anchors was added to the CNN to handle smaller boxes.
	\item The extra output for global image classification with one of the classes ('No Lung Opacity / Not Normal', 'Normal', 'Lung Opacity') was added to the model. This output was not used directly to classify the images, however, making the model predict the other related function improved the result.
    \item We added dropout~\cite{srivastava2014dropout} to the global classification output to reduce overfitting. In addition to extra regularisation, it helped to achieve the optimal classification and regression results around the same epoch.
\end{enumerate}

\subsection{Model training}
The training dataset included data for $25684$ patients and the test set had data for $1000$ patients. We used a range of base models pre-trained on ImageNet dataset~\cite{imagenet}. The models without pre-train on the ImageNet performed well on classification, but worse on regression task. The following hyper-parameters were used for all training experiments (Table 2):
\begin{table}[htb]
  \begin{center}
  \begin{tabular}{|l|c|}
  \hline
  Parameter & Description \\
  \hline\hline
      Optimizer & Adam  \\
      Initial learning rate & 1e-5  \\
      Learning rate scheduler& ReduceLROnPlateau \\
      Patience & 4 \\
      Image size & $512\times512$ \\
  \hline
  \end{tabular}
  \end{center}
  \label{table:params}
\caption{Common models hyper-parameters.}
\end{table}
As the training dataset was reasonably balanced (see Table 1), there was no need for extra balancing techniques. For learning rate scheduler we used available in Pytorch \texttt{ReduceLROnPlateau} with the patience of $4$ and learning rate decrease factor of $0.2$. The losses of whole image classification, individual boxes classification and anchors regression were combined with weights and used as a total loss.

\subsection{Model encoders}
\begin{figure}[htb]
\begin{center}
\includegraphics[width=1\linewidth]{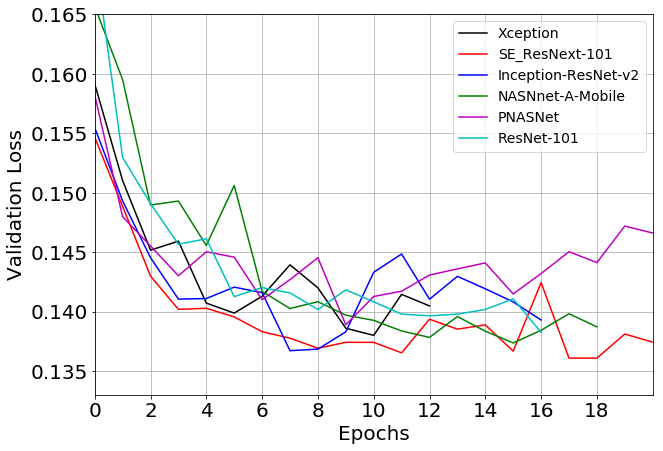}
\end{center}
   \caption{Evolution of the validation loss during training for the RetinaNet model with various encoders.}
\label{fig:train}
\end{figure}

A number of different encoder architectures has been tested: Xception~\cite{xception}, NASNet-A-Mobile~\cite{nasnet}, ResNet-34, -50, -101~\cite{resnet}, SE-ResNext-50, -101~\cite{Hu_2018_CVPR}, and DualPathNet-92~\cite{dpn}, Inception-ResNet-v2~\cite{inc}, PNASNet-5-Large~\cite{pnasnet}. In order to enable reasonably fast experiments and model iterations, we considered architectures with good trade-offs between accuracy and complexity/parameters number, and hence training time~\cite{Bianco2018}. In this regard, VGG nets~\cite{vgg} and MobileNets~\cite{mobilenet} do not provide optimal accuracy on ImageNet dataset~\cite{imagenet}, while SeNet-154~\cite{Hu_2018_CVPR} and NasNet-A-Large~\cite{nasnet} have the largest number of parameters and require the most floating-point operations. Fig.~\ref{fig:train} shows validation loss during training for various encoders used in the RetinaNet SSD. The SE-ResNext architectures demonstrated optimal performance on this dataset with a good trade-off between accuracy and complexity~\cite{Bianco2018}.

\subsection{Multi-task learning}
The extra output for global image classification with one of the classes ('No Lung Opacity / Not Normal', 'Normal', 'Lung Opacity') was added to the model. The total loss was combined of this global classification output with regression loss and individual boxes classification loss.

\begin{figure}[htb]
\begin{center}
\includegraphics[width=1\linewidth]{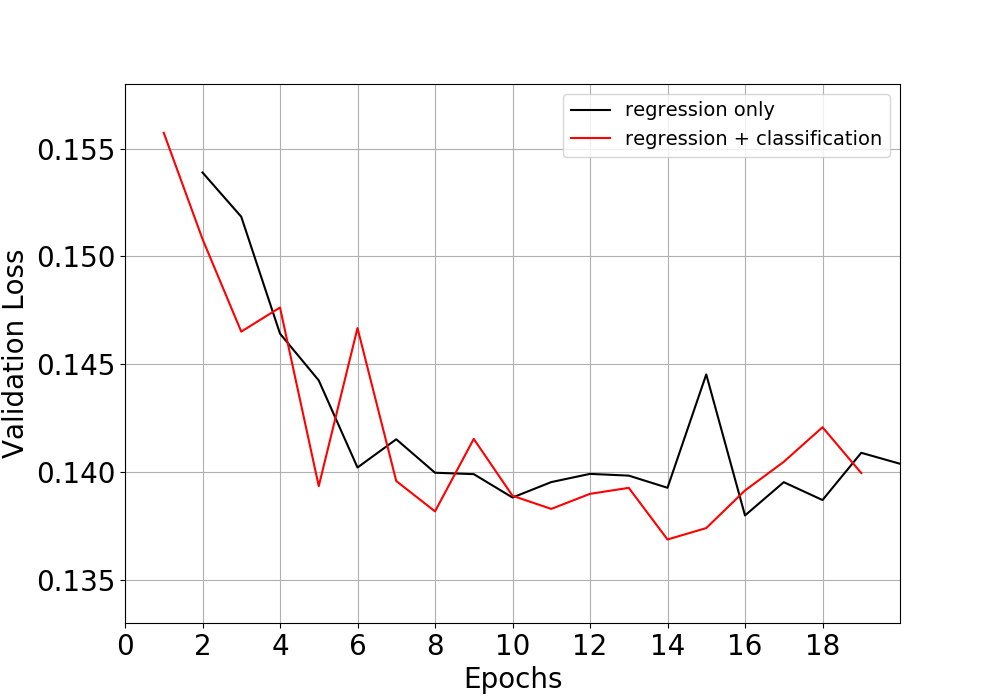}
\end{center}
   \caption{Evolution of the validation loss during training of RetinaNet with SE-ResNext-101 encoder with (red) and without (black) multi-task learning.}
\label{fig:losses}
\end{figure}

For an ablation study, we trained the RetinaNet model with the SE-ResNext-101 encoder and fixed augmentations with and without the global classification output. The training dynamics is shown in Fig.~\ref{fig:losses}. The output of global classification was not used directly to classify the images, however, making the model predict the other related function improved the result compared to training the regression-only output of the model.

\begin{figure}[htb]
\begin{center}
\includegraphics[width=1\linewidth]{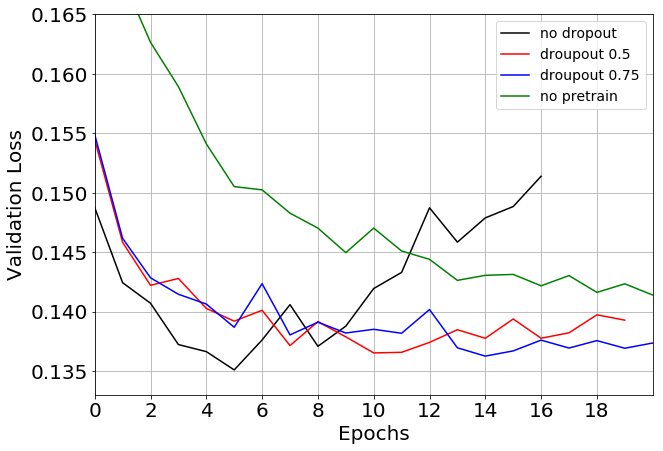}
\end{center}
   \caption{Evolution of the validation loss during training for different versions of RetinaNet with SE-ResNext-101 encoders.}
\label{fig:senets}
\end{figure}

As the classification output overfits faster than the detected anchors' positions/size regression, we added a dropout for the global image classification output. Besides regularization, it helped to achieve the optimal classification and regression results around the same epoch. Various dropout probabilities have been tested. Fig.~\ref{fig:senets} shows examples of training curves for SE-ResNext-101 with different dropouts and pre-training. Without a pre-training, the models took much longer to converge. RetinaNet SSD with the SE-ResNext-101 encoder pre-trained in Imagenet and with dropouts of $0.5$ and $0.75$ for the global classification output showed the best test metrics on this dataset.

\section{Images preprocessing and augmentations}
\begin{figure*}
\begin{center}
    \includegraphics[width=1\linewidth]{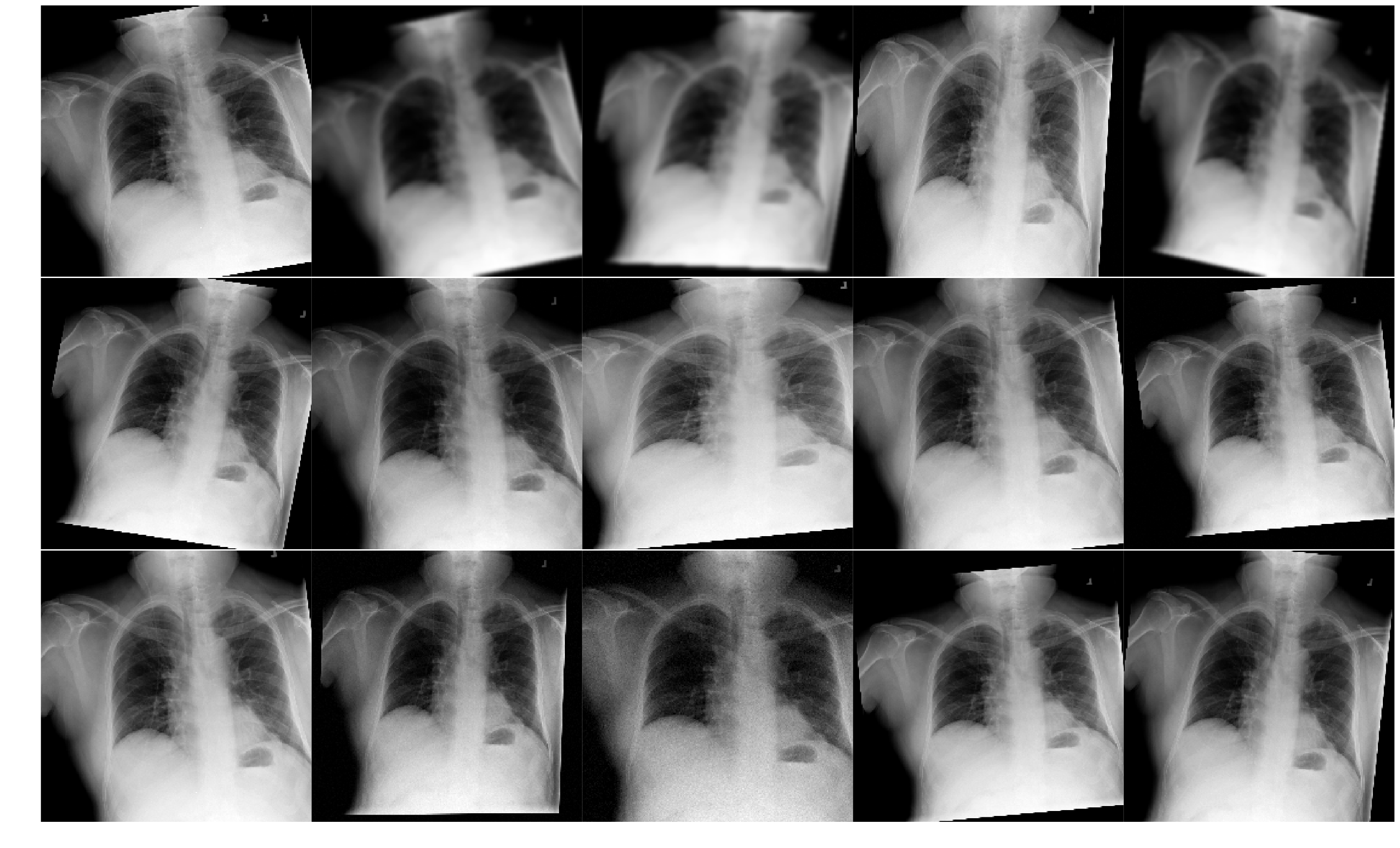}
\end{center}
   \caption{The example of a patient chest X-ray scan with heavy augmentations and rotations.}
\label{fig:aug}
\end{figure*}

The original images were scaled to the $512\times512$px resolution. The 256 resolution yielded a degradation of the results, while the full original resolution (typically, over $2000\times2000$px) was not practical with heavier base models. Since the original challenge dataset is not very large the following images augmentations were employed to reduce overfitting:
\begin{itemize}
	\item mild rotations (up to 6 degrees)
	\item shift, scale, shear
	\item horizontal flip
    \item for some images random level of blur, noise and gamma changes
    \item a limited the amount of brightness / gamma augmentations
\end{itemize}
An example of a patient X-ray scan with heavy augmentations is shown in Fig.~\ref{fig:aug}.

\subsection{Ablation study}
\begin{figure}[htb]
\begin{center}
    \includegraphics[width=0.7\linewidth]{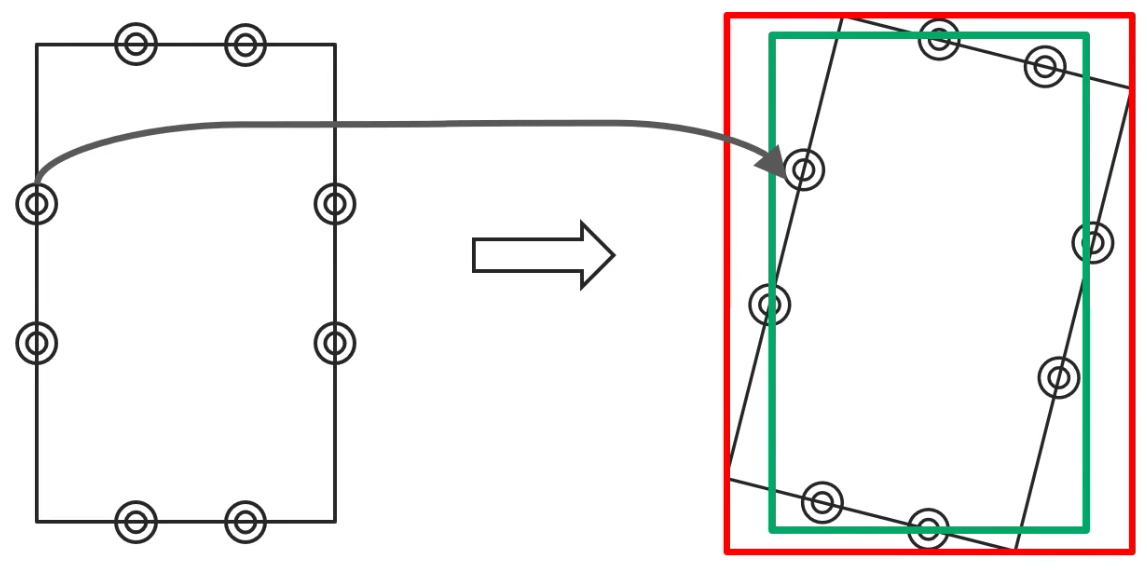}
\end{center}
   \caption{The diagram illustrating custom rotation of bounding boxes.}
\label{fig:rotate}
\end{figure}

To examine experimentally the effect of image augmentations, we conducted an ablation study with different augmentation sets. In this ablation study, we ran training sessions on the same model with fixed hyper-parameters and only changed the sets of image augmentations.
We used the following augmentation sets:
\begin{enumerate}
\item No augmentations: after resizing and normalisation, no changes were applied to the images
\item Light augmentations: affine and perspective changes (scale=0.1, shear=2.5), and rotations (angle=5.0)
\item Heavy augmentations: random horizontal flips, affine and perspective changes (scale=0.15, shear=4.0), rotations (angle=6.0), occasional Gaussian noise, Gaussian blur, and additive noise
\item Heavy augmentations without rotation: heavy augmentations described above without rotations
\item Heavy augmentations with custom rotation: heavy augmentations described above with mild rotations of 6 deg, customised as shown in Fig.~\ref{fig:rotate}
\end{enumerate}

\begin{figure}[htb]
\begin{center}
    \includegraphics[width=0.9\linewidth]{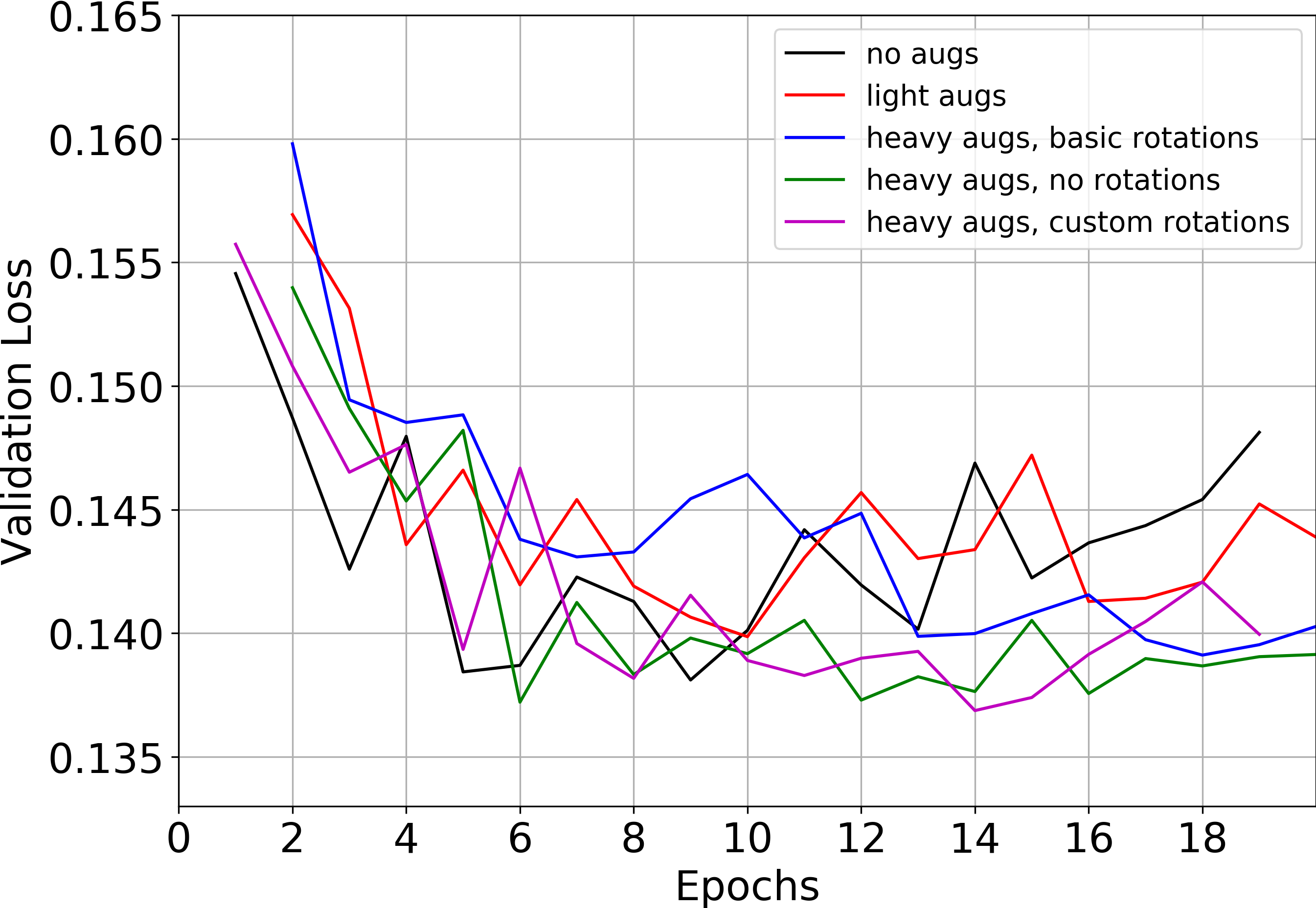}
\end{center}
   \caption{Evolution of the validation loss during training for different sets of augmentations.}
\label{fig:augs}
\end{figure}

The dynamics of the training with different sets of augmentations is shown in Fig.~\ref{fig:augs}.

The results for all experiments are presented in Table 3.
\begin{table}[htb]
\begin{center}
\begin{tabular}{|l|c|c|}
\hline
Augmentations & Best validation mAP \\
\hline\hline
no augmentations & 0.246127   \\
light augmentations & 0.254429  \\
heavy augmentations & 0.250230  \\
heavy augmentations \\ custom rotation &0.255617 \\
heavy augmentations, \\no rotation & 0.260971  \\
\hline
\end{tabular}
\end{center}
\label{table:augs}
\caption{Pneumonia detection mean average precision results achieved with various augmentations sets on validation.}
\end{table}

Without enough image augmentations the model showed signs of overfitting when the validation loss stopped improving (see Fig.~\ref{fig:augs}). With light and heavy augmentations, the same model showed better validation loss and mAP scores. The image rotations had a measurable effect on the results, as the rotation of the bounding boxes around corners modifies the original annotated regions significantly. To reduce the impact of the rotation on bounding box sizes, instead of rotating the corners we rotated two points at each edge, at 1/3 and 2/3 edge length from the corner (8 points in total), and calculated the new bounding box as min/max of the rotated points, as illustrated in Fig.~\ref{fig:rotate}. We tested the same model with usual rotation, custom rotation and no rotation at all. The custom rotation improved the results, but the heavy augmentations without any rotation gave the best metrics on the validation.

\section{Postprocessing}
There was a difference in train and test the labelling process of the dataset provided. The train set was labelled by a single expert, while the test set was labelled by three independent radiologists and the intersection of their labels was used for the ground truth~\cite{shih2019augmenting}. This yielded a smaller labelled bounding box size, especially in complicated cases. This process can be simulated using outputs from $4$ folds and/or predictions from multiple checkpoints. The $20$ percentile was used instead of the mean output of anchor sizes, and then it was reduced even more, proportionally to the difference between $80$ and $20$ percentiles for individual models (with the scale of $1.6$ optimised as a hyper-parameter).

\begin{figure}[htb]
\begin{center}
    \includegraphics[width=1\linewidth]{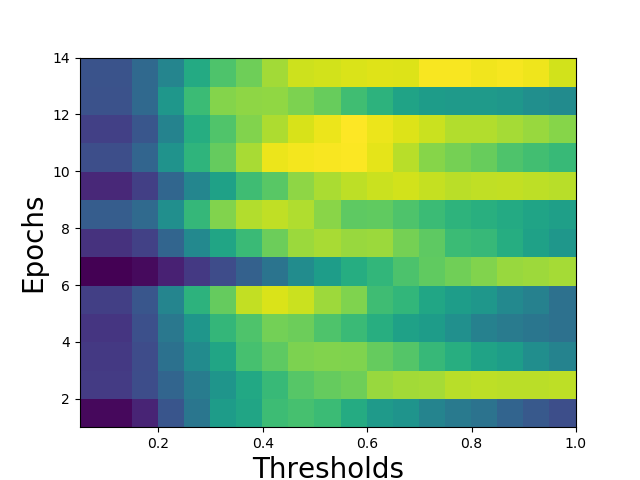}
\end{center}
   \caption{The validation mAP metric versus epochs and NMS thresholds.}
\label{fig:maps}
\end{figure}

The optimal threshold for the non-maximum suppression (NMS) algorithm was also different for the train and test sets due to different labelling process. The test set true labels were available after the challenge. The NMS thresholds had a dramatic impact on the mAP metric values. Fig.~\ref{fig:maps} shows the validation mAP metrics evolution for different training epochs and NMS thresholds. The optimal NMS thresholds on validation set varied significantly from epoch to epoch with the optimum between $0.45$ and $1$ depending on the model.

The other approach is re-scaling the predicted boxes sizes for the test set to $87.5\%$ of the original sizes to reflect the difference between test and train set labelling process. The coefficient of $87.5\%$ was chosen to approximately match the sizes to the previous approach. These differences between the train and test sets reflect differences in the annotation process for these datasets, with a consensus of expert radiologists used as ground truth in the test sets.

\section{Results}
\begin{figure}[htb]
\begin{center}
    \includegraphics[width=1\linewidth]{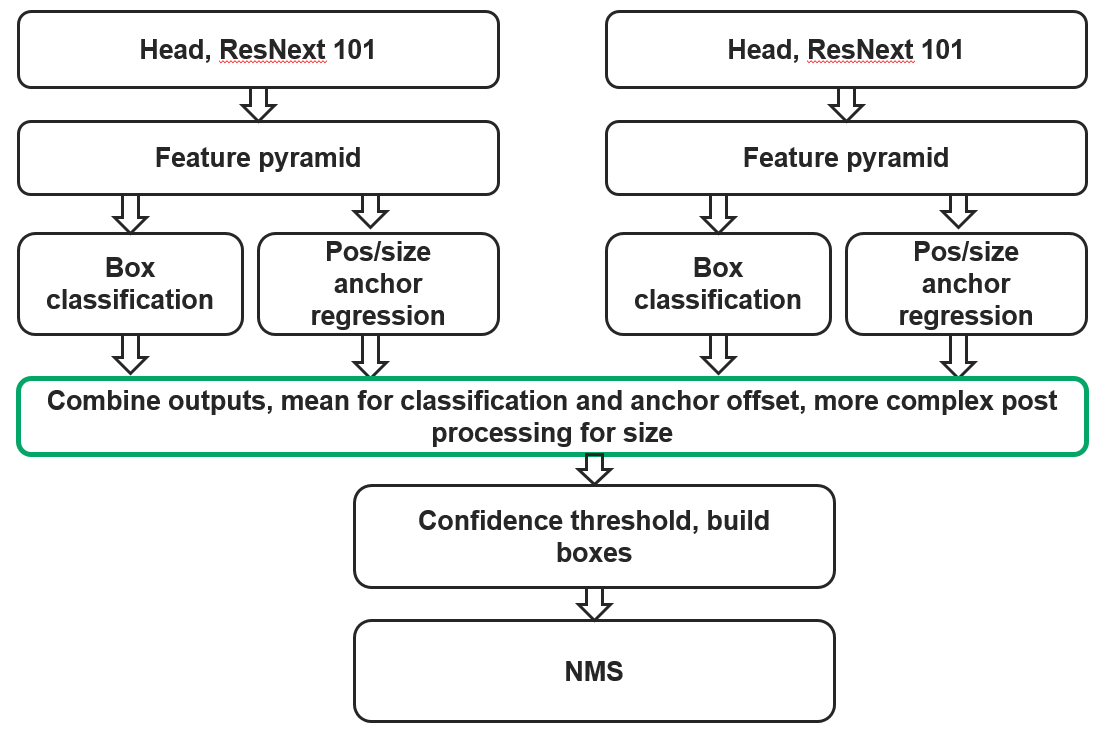}
\end{center}
   \caption{The diagram of the same model ensemble technique.}
\label{fig:model}
\end{figure}

The results of detection models can change significantly between epochs and depend largely on thresholds. Therefore, it is beneficial to ensemble models from different checkpoints to achieve a more stable and reliable solution. The outputs from the same model for $4$ cross-validation folds and several checkpoints were combined before applying NMS algorithms and optimizing thresholds (see the diagram of the ensemble in Fig.~\ref{fig:model}.

The final top results of the challenge are shown in Table 4.
\begin{table}[htb]
\begin{center}
\begin{tabular}{|l|c|c|}
\hline
Team Name & Test set, mAP \\
\hline\hline
Ian Pan and Alexandre Cadrin-Chênevert & 0.25475   \\
Dmytro Poplavskiy & 0.24781 \\
Phillip Cheng & 0.23908 \\
\hline
\end{tabular}
\end{center}
\label{table:results}
\caption{The final leader board results in Pneumonia detection challenge showing mAP metric calculated on the private test set.}
\end{table}

The method described in this paper took second place in the challenge. The model was based on RetineNet SSD with Se-ResNext101 encoders pre-trained on ImageNet dataset, heavy augmentations with custom rotation as described in Section 6, multi-task learning with global classification output (see Section 5) and postprocessing as in Section 7. For the final ensemble, the outputs from the same model for $4$ cross-validation folds and several checkpoints were combined before applying NMS algorithms (as shown in Fig.~\ref{fig:model}). The postprocessing with re-scaling predictions was applied to compensate for the difference between the train and test sets labelling processes.

\section{Discussion}
The other winner's solutions were also based on the ensemble of RetinaNet models with various inputs and encoders\cite{Pan2020}.
Remarkably, all top teams made similar discoveries regarding the differences between the training and test sets. All three teams found
that lowering threshold for the NMS algorithm for the test predictions compared to the validation set improved the test set scores.

In addition, systematic size reductions of the predicted bounding boxes have been also applied by the other winning teams \cite{Pan2020}. These difference between the train and test set reflect differences in the datasets labelling process. The train set was labelled by a single expert, while the test set was labelled by three independent radiologists and the intersection of their labels was used for the ground truth.

\section{Conclusions}
In this paper, we propose a simple and effective algorithm for the localization of lung opacities regions. The model was based on single-shot detector RetinaNet with Se-ResNext101 encoders, pre-trained on ImageNet dataset. The number of improvements was implemented to increase the accuracy of the model. In particular, the global classification output added to the model, heavy augmentations were applied to the data, the ensemble of $4$ folds and several checkpoints was unitised to generalise the model. Ablation studies have shown the improvements by the proposed approaches for the model accuracy. This method purposely does not involve test-time augmentation and provides a good trade-off between accuracy and resources. The reported method achieved one of the best results in the Radiological Society of North America (RSNA) Pneumonia Detection Challenge.

\section{Acknowledgements}
We thank the National Institutes for Health Clinical Center for providing the chest X-ray images used in the competition, Kaggle, Inc. for hosting the challenge. The authors thank Google Cloud Platform and Dutch internet service provider HOSTKEY B.V. (hostkey.com) for access to GPU servers and technical assistance. We also acknowledge the Radiological Society of North America, the Society of Thoracic Radiology, and Kaggle, Inc. for annotating the images and organizing the competition. The authors thank the Open Data Science community (ods.ai) for useful suggestions. A.K.K. thanks Xin Rong of the University of Michigan for the donation of the Titan X NVIDIA GPU.

{\small
\bibliographystyle{ieee_fullname}
\bibliography{rsnabib}
}

\end{document}